\documentclass[sigconf]{acmart}
\usepackage{color}
\usepackage{makecell}
\usepackage{multirow}
\usepackage{enumitem}

\usepackage{amsmath}
\usepackage{mdframed}
\usepackage{xcolor}
\usepackage{hyperref}
\usepackage{subcaption}
\usepackage{bm}

\graphicspath{{figures/}}
\newmdenv[
    topline=true,
    bottomline=true,
    leftline=true,
    rightline=true,
    linecolor=blue,
    roundcorner=20pt,
    skipbelow=8pt,
    skipabove=10pt,
]{reviewbox}

\AtBeginDocument{%
  \providecommand\BibTeX{{%
    \normalfont B\kern-0.5em{\scshape i\kern-0.25em b}\kern-0.8em\TeX}}}

\copyrightyear{2025} \acmYear{2025} 
\acmConference[UbiComp Companion '25]{Companion of the 2025 ACM International Joint Conference on Pervasive and Ubiquitous Computing}{October 12--16, 2025}{Espoo, Finland} \acmBooktitle{Companion of the 2025 ACM International Joint Conference on Pervasive and Ubiquitous Computing (UbiComp Companion '25), October 12--16, 2025, Espoo, Finland}\acmDOI{10.1145/3714394.3756175} \acmISBN{979-8-4007-1477-1/2025/10}

\begin{document}

\title{Contact Sensors to Remote Cameras: Quantifying Cardiorespiratory Coupling in High-Altitude Exercise Recovery}



\begin{abstract}
    Cardiorespiratory coupling (CRC) captures the dynamic interaction between the cardiac and respiratory systems—an interaction strengthened by physical exercise and linked to improved physiological function. We examined CRC at high altitude in two states, rest and post-exercise recovery, and found significant differences (p $\leq$ 0.05). Quantitative analysis revealed that recovery involved more frequent yet less stable episodes of synchronization between respiration and pulse. Furthermore, we explored the feasibility of non-contact CRC measurement with remote photoplethysmography (rPPG), observing a strong correlation with oximeter-based metrics (Pearson r = 0.96). These findings highlight the potential of CRC as a sensitive marker for autonomic regulation and its future application in contactless monitoring. Source code is available at GitHub: \url{https://github.com/McJackTang/CRC}.

\end{abstract}




\begin{teaserfigure}
    \centering
    \includegraphics[width=1\linewidth]{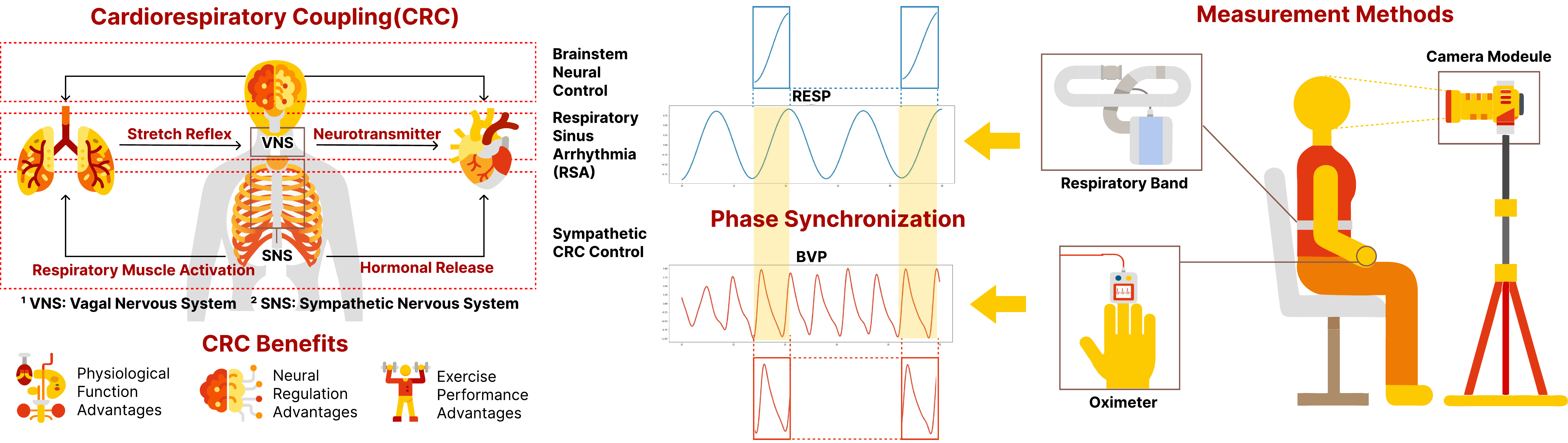}
    \caption{\textbf{Framework of cardiorespiratory coupling: physiological mechanisms and measurement methods.} Cardiorespiratory coupling is driven by autonomic (vagal and sympathetic) regulation via neurotransmitters and hormones, producing respiratory sinus arrhythmia and phase locking between pulse and breath; camera‑based rPPG enables promising non‑contact assessment.}
    \label{fig:overview}
\end{teaserfigure}



\author{Jiankai Tang}
\orcid{0009-0009-5388-4552}
\email{tjk24@mails.tsinghua.edu.cn}
\author{Meng Kang}
\orcid{0009-0009-0796-5132}
\author{Yiru Zhang}
\orcid{0009-0001-8895-0902}
\author{Kegang Wang}
\orcid{0000-0003-3469-0786}
\affiliation{%
  \institution{Tsinghua University}
  \country{China}
}
\author{Daniel Mcduff}
\orcid{0000-0001-7313-0082}
\author{Xin Liu}
\orcid{0000-0002-9279-5386}
\email{xliu0@cs.washington.edu}
\affiliation{%
  \institution{University of Washington}
  \country{USA}
}
\author{Yuanchun Shi}
\orcid{0000-0003-2273-6927}
\author{Yuntao Wang}
\authornote{Corresponding author}
\email{yuntaowang@tsinghua.edu.cn}
\orcid{0000-0002-4249-8893}
\affiliation{%
  \institution{Tsinghua University}
  \country{China}
}



\maketitle

\section{Introduction}

\begin{figure*}[t]
    \centering
    \includegraphics[width=1\linewidth]{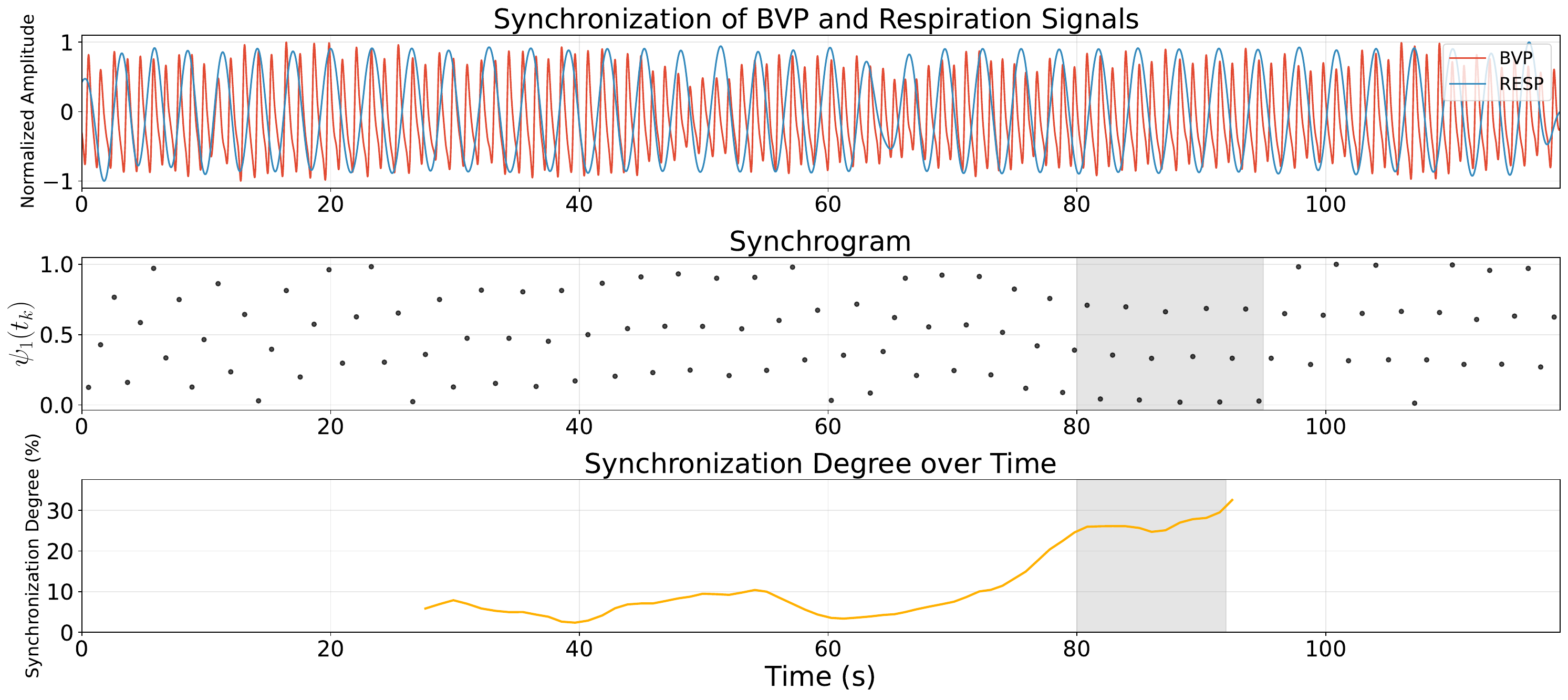}
        \vspace{-0.8cm}
    \caption{\textbf{Visualization of cardiorespiratory waveforms and synchronization.} This figure shows normalized BVP and RESP waveforms, a synchrogram illustrating n:1 phase synchronization—where n heartbeats align at consistent phases across one respiratory cycle, and the resulting synchronization degree over time.}
\vspace{-0.5cm}
    \label{fig:vis-crc}
\end{figure*}

Cardiorespiratory coupling (CRC) reflects the dynamic interaction between the cardiac and respiratory systems, mediated through shared central and peripheral neural pathways~\cite{perry2019physical}. CRC has been observed in both healthy individuals and clinical populations, indicating its fundamental role in autonomic regulation \cite{garcia2013cardiorespiratory}. This coupling is evident in phenomena such as respiratory sinus arrhythmias (RSA), where the heart rate increases during inspiration and decreases during expiration~\cite{ma2024sleep}.

CRC is closely related to physiological function, with higher synchronization generally indicating better cardiorespiratory fitness and recovery capacity. Abnormal CRC patterns can signal issues such as obstructive sleep apnea and other cardiopulmonary disorders~\cite{kabir2010cardiorespiratory}. Reliable quantification of CRC ratio and degree is therefore critical for clinical research and health monitoring \cite{huang2021cardiorespiratory}.

CRC is commonly assessed with synchrogram analysis, which captures the phase relationship between heartbeats and respiration. Building on prior work \cite{schafer1998heartbeat}, we automated the extraction of long-term CRC ratio and degree from cardiorespiratory phase signals, while using synchrograms as a visual tool to illustrate the n:m phase locking patterns, and further examined respiratory-phase alignment within individual cardiac cycles. Comparing physiological states, we observed significant CRC changes (p $\leq$  0.05) during post-exercise recovery versus rest at high altitude.

Traditional CRC assessments rely on contact sensors (ECG, PPG, respiratory bands), limiting everyday practicality \cite{wu2010cardiorespiratory}. We therefore evaluated remote photoplethysmography (rPPG) as a non-contact alternative. rPPG estimates pulse waveforms from subtle skin-reflectance changes, enabling unobtrusive deployment. CRC metrics derived from rPPG exhibited a strong correlation with those from contact devices (Pearson r = 0.956), demonstrating rPPG’s promise for remote monitoring. To the best of our knowledge, this is the first study to characterize CRC during exercise recovery at high altitude using both contact and non-contact modalities.

The key contributions of this paper include:
     
\begin{itemize}
\vspace{-0.3cm}
    \item We demonstrated significant CRC modulation (p $\leq$ 0.05) across rest and post-exercise recovery in a high-altitude environment.
    \item We systematically quantified long-term CRC ratio and degree and revealed short-cycle synchronization patterns between respiratory and pulse phases.
    \item We explored that CRC measured via rPPG closely matches contact-based measurements (Pearson r = 0.956), highlighting a viable path toward ubiquitous, non-contact cardiorespiratory monitoring.
         \vspace{-0.3cm}

\end{itemize}

\section{Related Work}
\subsection{Cardiorespiratory Coupling Measurement}

 \begin{figure*}[t]
    \centering
    \includegraphics[width=1\linewidth]{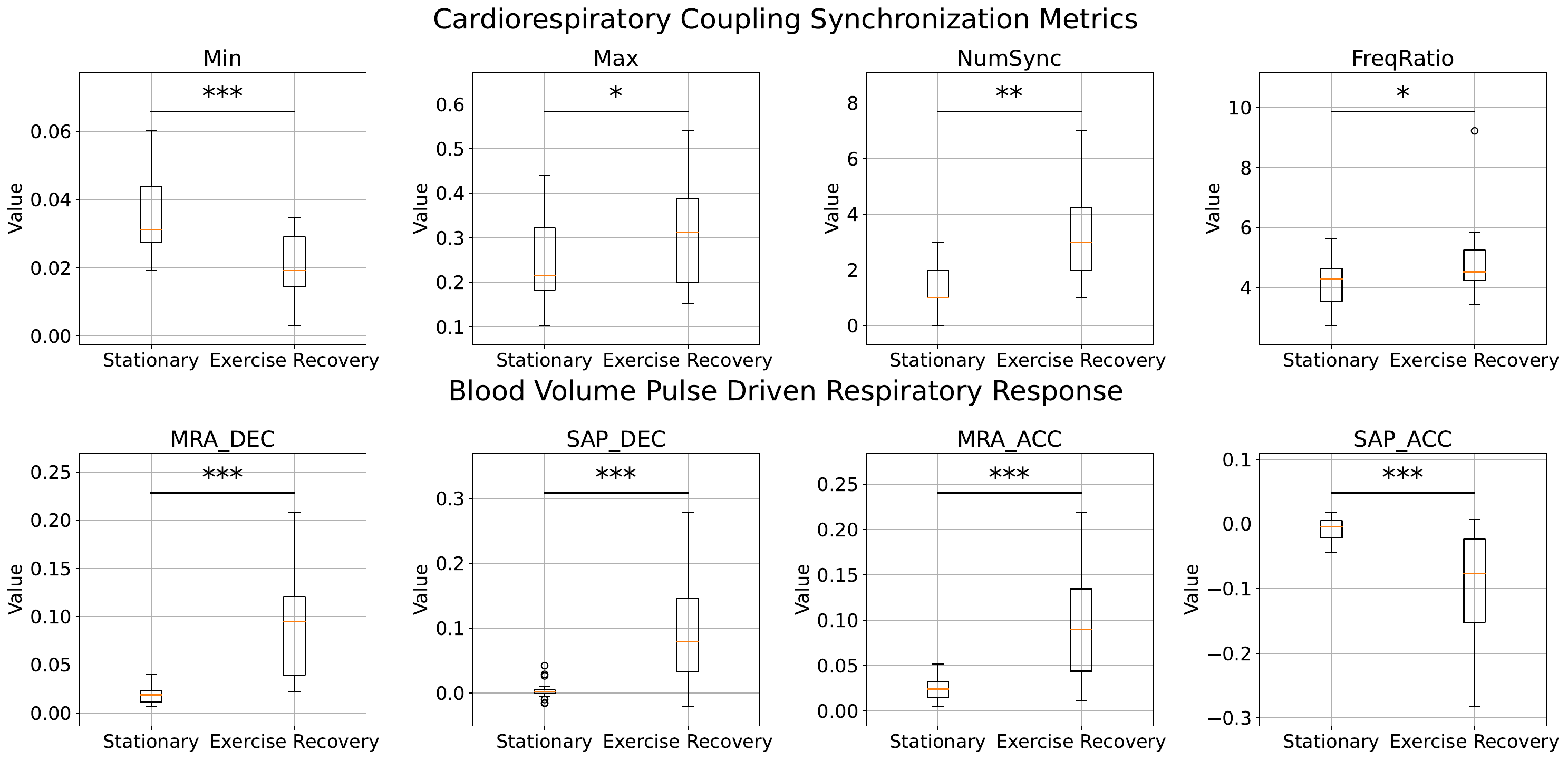}
        \vspace{-0.8cm}
    \caption{Box plots of cardiorespiratory coupling long-term metrics (Min, Max, NumSync, FreqRatio) and short-term metrics (MRA, SAP) across different physiological states.}
\vspace{-0.5cm}
    \label{fig:box}
\end{figure*}


Early studies~\cite{schafer1998heartbeat} reported the phenomenon of cardiorespiratory phase synchronization (CRPS) under free-running conditions and proposed an analytical approach to detect synchronization periods of various \textit{n:m} ratios. This provided evidence for a previously unrecognized form of cardiorespiratory interaction. Research demonstrated that CRPS could last up to 20 minutes during spontaneous breathing at rest. 
Subsequent research~\cite{wu2010cardiorespiratory} explored CRC during normal rest and meditation, showing that CRC can be enhanced during meditation and is most apparent under conditions of low cognitive load. Further investigations~\cite{sola2018cardiorespiratory} later found that psychological stimulation can also increase CRC when aerobic exercise is not feasible. However, CRC under high-altitude exercise recovery has not been widely investigated.


In the early study~\cite{schafer1998heartbeat}, ECG and nasal airflow (measured with thermistors) were recorded simultaneously. The synchronization analysis was based on phase-locking theory for chaotic oscillators, which allowed analysis of irregular and nonstationary bivariate data. By examining the respiratory phase at each heartbeat time $t_{k}$ and identifying the preferred phases of \textit{n} beats over \textit{m} respiratory cycles, the evolution of the relative phases was visualized as \textit{n} horizontal lines in a synchrogram. 


With the development of signal processing, more recent studies have applied various linear and nonlinear methods for CRC analysis, such as joint symbolic analysis, squared coherence, Granger causality, transfer entropy, and principal component analysis~\cite{abreu2023significance}. Each method captures different CRC properties, such as directionality or nonlinear characteristics, and the choice of method depends on the research context. Notably, most prior studies have not used non-contact measurement approaches for CRC.




\subsection{Application of Cardiorespiratory Coupling}

Cardiorespiratory coupling (CRC) reflects the dynamic interaction between the cardiovascular and respiratory systems, and has emerged as a sensitive marker of autonomic regulation. Clinical studies have linked CRC to multiple chronic conditions, including obstructive sleep apnea~\cite{kabir2010cardiorespiratory}, type 2 diabetes mellitus~\cite{da2023cardiorespiratory}, and periodic breathing patterns observed in elderly individuals~\cite{giraldo2022cardiorespiratory}, which may arise from age-related changes or comorbidities. CRC-based metrics have been shown to outperform traditional vital signs and heart rate variability in predicting extubation outcomes in ICU patients~\cite{armanac2021cardiopulmonary}. These findings underscore CRC’s value in early diagnosis and clinical decision-making, offering a non-invasive approach to detect autonomic dysfunction even in asymptomatic individuals~\cite{garcia2013cardiorespiratory}.

Beyond physical health, CRC has shown relevance in neurological and psychiatric research. Studies have found that schizophrenia patients exhibit weakened CRC strength and altered synchronization directions, indicative of autonomic imbalance~\cite{schulz2014analyses}. Interestingly, similar CRC alterations have been observed in first-degree relatives~\cite{berger2010reduced}, suggesting potential heritability. In affective computing, incorporating CRC features has significantly improved the accuracy of emotion recognition systems~\cite{valenza2013improving}. These results highlight the broader utility of CRC as a physiological marker for mental state monitoring and early detection of neuropsychiatric disorders.

In recent years, CRC has gained attention in sports medicine due to its sensitivity to physical activity and training load~\cite{abreu2023significance}. CRC metrics can provide insight into cardiovascular adaptability, fatigue, and overtraining risks, helping optimize performance. For example, CRC strength correlates positively with exercise capacity in COPD patients~\cite{huang2021cardiorespiratory}. However, despite increasing interest, CRC remains under-investigated in hypoxic and high-altitude exercise scenarios, where autonomic regulation is significantly challenged. These settings are highly relevant for athletes, military personnel, and individuals exposed to altitude-related stress. Future research should address this gap, as CRC may offer a valuable non-invasive tool for evaluating physiological resilience and training readiness in extreme environments.

\subsection{Remote Cardiorespiratory Sensing}

Remote physiological sensing aims to extract vital signs such as heart rate (HR), respiration, and blood oxygen saturation (SpO\textsubscript{2}) without physical contact, typically using video cameras. The core principle involves analyzing subtle color changes in facial skin regions caused by blood volume fluctuations, a phenomenon known as remote photoplethysmography (rPPG)~\cite{liu2024rppg}. These periodic signals can be extracted using signal processing or deep learning techniques to infer HR~\cite{wang2025memory}. Similarly, respiratory information can be derived either from rPPG amplitude modulation or from subtle head and chest movements, while SpO\textsubscript{2} estimation relies on modeling the differential absorption of light at multiple wavelengths, typically inferred through RGB channels in standard video~\cite{verkruysse2008remote,tang2025camera}.

Recent studies have demonstrated successful implementation of these principles. For example, TS-CAN~\cite{liu2020multi} introduced a dual-branch temporal shift network to simultaneously estimate rPPG and respiration waveforms. MultiPhysNet~\cite{liu2024summit} and FusionPhys~\cite{ma2025non} further extended these capabilities to high-altitude environments, achieving robust estimation of both PPG signals, respiration waveforms and SpO\textsubscript{2} from facial video under challenging conditions. Other works~\cite{tang2023mmpd,liu2024spiking} have validated the accuracy and reliability of remote HR sensing across varied scenarios. While these studies demonstrate the growing feasibility of non-contact cardiorespiratory monitoring, few have explored the estimation of CRC from facial videos.

\section{Method}

\subsection{Dataset}
~\label{sec:dataset}

We utilized the SUMS dataset~\cite{liu2024summit} for our CRC analysis under high-altitude conditions. This dataset includes synchronized physiological signals collected from ten participants (age $25.5 \pm 1.44$ years) on the Qinghai Plateau, comprising 80 recording periods of 2-3 minutes each. The data includes pulse waveforms captured with an FDA-approved CMS50E+ oximeter (20 Hz), respiratory signals from an HKH-11C sensor (50 Hz), and synchronised facial videos. The experimental protocol alternates between stationary phases (stages 1 and 3) and exercise recovery phases (stages 2 and 4), creating a comprehensive dataset that captures cardiorespiratory dynamics across different physiological states. Due to high‑altitude exercise conditions, heart and respiratory rates accelerate and blood oxygen levels decrease, posing unique challenges and yielding more pronounced CRC differences than in lowland settings. This multimodal and precisely synchronized data is particularly valuable for studying CRC in challenging high-altitude environments.

\begin{figure}
    \centering
    \includegraphics[width=1\linewidth]{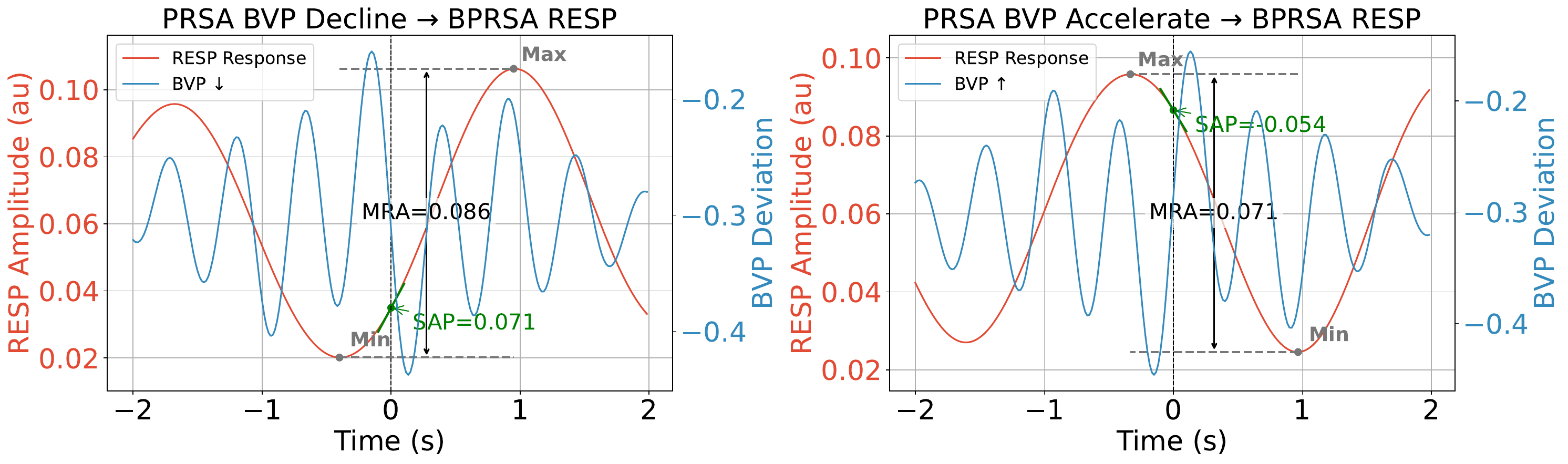}
        \vspace{-0.8cm}
    \caption{Visualization of PRSA and BPRSA for extracting respiratory features from decline and acceleration BVP phase.}
\vspace{-0.5cm}
    \label{fig:bprsa}
\end{figure}

\subsection{CRC Degree Calculation}
~\label{sec:crc degree}

We used the synchrogram method to analyze cardiorespiratory phase synchronization (CRPS) under high-altitude conditions~\cite{schafer1998heartbeat}. First, we detected peaks in the Blood Volume Pulse (BVP) signal to approximate heartbeat times ($t_{k}$). The instantaneous phase of the respiratory signal was then obtained using the Hilbert transform. At each heartbeat time $t_{k}$, we recorded the corresponding respiratory phase. The normalized relative phase over \textit{m} respiratory cycles was calculated as Eqn.~\ref{eq:psi_mod}:

\begin{equation}\label{eq:psi_mod}
\psi_{m}(t_{k})
= \frac{1}{2\pi}\bigl[\phi_{r}(t_{k}) \bmod (2\pi m)\bigr]
\end{equation}


 In Figure~\ref{fig:vis-crc},the presence of n horizontal lines in the synchrogram, which plots the respiratory phase $\psi_{m}(t_{k})$ at each heartbeat, reflects an \textit{n:m} phase synchronization. The pattern and regularity of the normalized relative phase $\psi_{m}(t_{k})$ reflect the existence and strength of synchronization between heartbeats and respiration. To quantify this relationship more precisely, we further transformed $\psi_{m}(t_{k})$ into $\Psi_{n,m}(t_k)$ as Eqn.~\ref{eq:psi_nm}:

\begin{equation}
\label{eq:psi_nm}
\Psi_{n,m}(t_k)
= \frac{2\pi}{m}\bigl\{\bigl[\psi_{m}(t_k)\cdot n\bigr]\bmod m\bigr\}.
\end{equation}

In this way, the \textit{n} horizontal lines in the synchrogram are combined into a single line, allowing us to quantify the degree of \textit{n:m} phase synchronization using the index $\gamma_{n,m}$ calculated from Eqn.~\ref{eq:gamma_nm}, which measures the stability of $\Psi_{n,m}(t_k)$~\cite{wu2010cardiorespiratory}. This approach enables a clear and quantitative assessment of cardiorespiratory coupling strength across different physiological states and time scales.

\begin{equation}\label{eq:gamma_nm}
\gamma_{n,m}
= \left\{\frac{1}{N}\sum_{k}\cos\bigl[\Psi_{n,m}(t_k)\bigr]\right\}^2
+ \left\{\frac{1}{N}\sum_{k}\sin\bigl[\Psi_{n,m}(t_k)\bigr]\right\}^2\ .
\end{equation}


where $N$ represents the number of heartbeats within the analysis window. The synchronization index $\gamma_{n,m}$ ranges from 0 (no synchronization) to 1 (complete synchronization). In this study, we set $N=50$ (25 heartbeats before and after each reference point). To identify the strongest synchronization at each heartbeat, we defined the synchronization degree at time $t_{k}$ as the maximum $\gamma_{n,m}(t_{k})$ across all tested $(n, m)$ pairs. For each subject and physiological state, we calculated the minimum (Min) and maximum (Max) synchronization indices, the number of synchronization episodes (NumSync), and the average heart-to-respiration rate ratio (FreqRatio) as evaluation metrics in Table ~\ref{table_1}. A synchronization episode was defined as a continuous segment where $\gamma_{n,m}$ remained above 0.1 for at least 5 seconds.

\subsection{BVP Driven Respiration Coupling }
~\label{sec:bprsa}


Based on the findings of respiration-driven heart rate analysis~\cite{hirsch1981respiratory}, we further applied the Bivariate Phase-Rectified Signal Averaging (BPRSA) method to quantify the coupling between pulse wave and respiratory signals~\cite{joshi2019cardiorespiratory}. BPRSA extends the traditional PRSA approach by identifying anchor points (APs) in the pulse wave (trigger signal) and then averaging segments of the respiration signal (target signal) centered around each AP, with a default of 4 seconds. This method helps reveal subtle, quasi-periodic interactions between the heart and lungs that may be masked by noise or nonstationary fluctuations. The BPRSA process is illustrated in Fig~\ref{fig:bprsa}.
 

Unlike the long-term (minutes) CRC metrics described in Section~\ref{sec:crc degree}, BPRSA provides an average characterization of short-term (4-second) cardiorespiratory coupling within individual pulse wave cycles. For each BPRSA waveform, we extracted three key features, as summarized in Table~\ref{table_2}:
\begin{itemize}
    \item \textbf{Maximum Respiratory Amplitude (MRA):} The difference between the peak and trough of the averaged respiratory waveform, reflecting the largest change in breathing volume.
    \item \textbf{Slope at the Anchor Point (SAP):} The instantaneous slope at the anchor point, where positive values indicate inhalation and negative values indicate exhalation.
\end{itemize}

These features provide a clear and quantitative description of short-term heart–lung interactions under different physiological conditions, complementing the long-term CRC evaluation and supporting the systematic assessment of CRC dynamics.

\subsection{TN-rPPG for Non-Contact Signal Extraction}

To extract non-contact pulse waveforms from facial videos, we employed a pretrained Temporal Normalization-based rPPG model (TN-rPPG)~\cite{wang2024plug}, which has demonstrated strong robustness in unconstrained environments. The TN-rPPG model leverages a temporal normalization module to suppress illumination and motion-induced noise, producing cleaner pulse waveforms from RGB videos. Unlike traditional frame-differencing approaches, this model operates on temporally normalized features, using detrending and RMS normalization to reduce low-frequency drift and highlight physiological periodicities. In our study, we used TN-rPPG in the inference-only mode to generate remote photoplethysmographic (rPPG) signals from facial videos in the SUMS dataset.

\section{Results and Findings}

\begin{figure}
    \centering
    \includegraphics[width=1\linewidth]{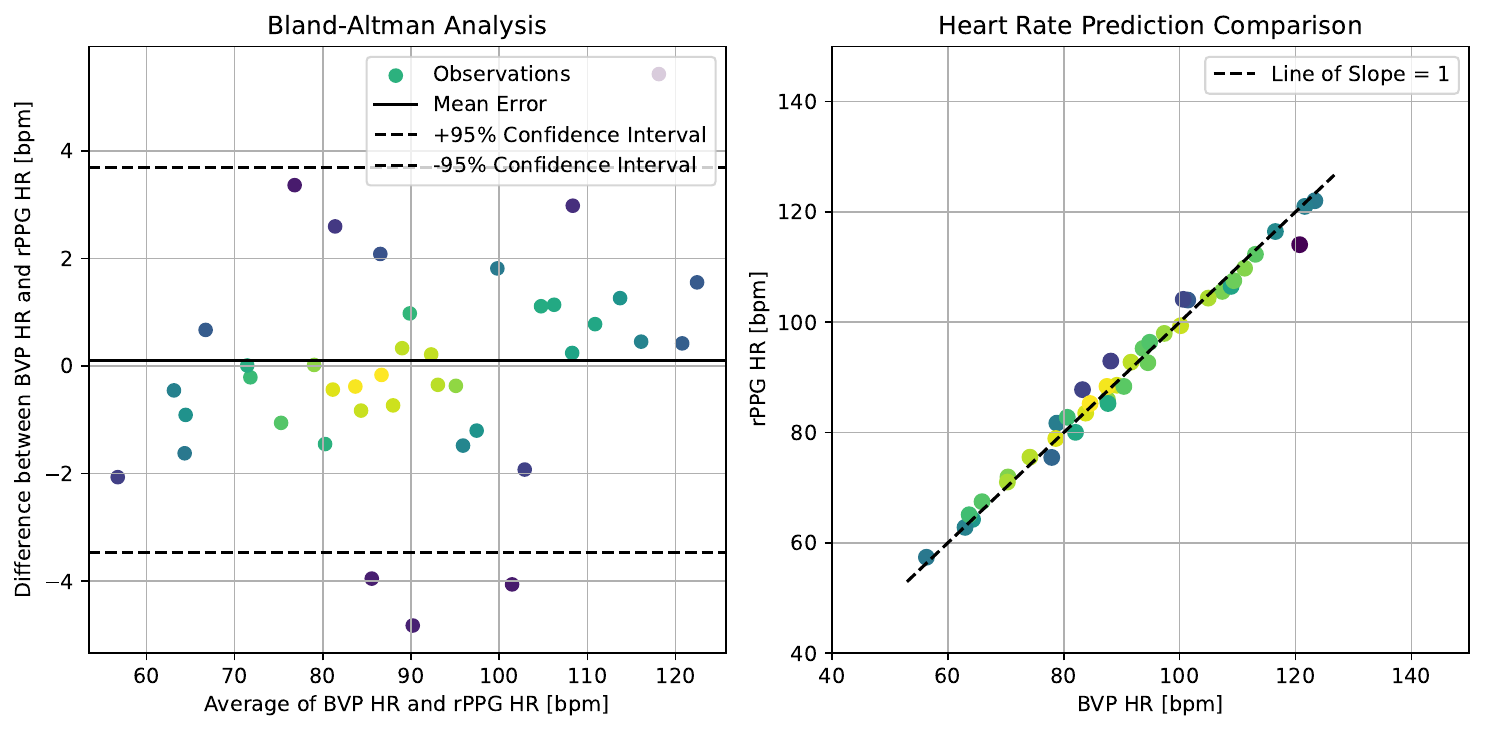}
        \vspace{-0.8cm}
    \caption{HR Prediction Compared with Ground Truth}
    \label{fig:HR}
    \vspace{-0.5cm}
\end{figure}

In this section, we discussed the main results and findings regarding CRC differences in Section ~\ref{sec: CRC varies}, synchronization phase in Section ~\ref{sec:resp_phase}, and remote CRC sensing ~\ref{sec:remote_rppg}.
CRC ratios and synchronization degrees were calculated from raw waveform data using synchrogram (Section ~\ref{sec:crc degree}), as illustrated in Fig ~\ref{fig:vis-crc}. The BPRSA method (Section ~\ref{sec:bprsa}) was applied to extract respiratory features from pulse waveforms, as shown in Fig ~\ref{fig:bprsa}. The significance testing was performed using the Wilcoxon signed-rank test, with significance levels of $p<0.05$(*), $p<0.01$(**), and $p<0.001$(***).

\begin{table}
    \centering
    \setlength{\tabcolsep}{3pt}
    \caption{Synchronization Metrics}
    \vspace{-0.3cm}
    \label{table_1}
    \resizebox{\columnwidth}{!}{
    \begin{tabular}{lcccc}
    \toprule
    State & Min (\%) & Max (\%) & NumSync & FreqRatio \\
    \midrule
    Stationary & $3.53 \pm 1.14$ & $25.17 \pm 9.20$ & $1.55\pm0.92$ & $4.15\pm0.79$ \\ 
    Exercise Recovery & $2.05 \pm 0.97$ & $31.69\pm12.29$ & $3.15\pm1.59$ & $4.82\pm1.20$ \\ 
    \bottomrule  
    \end{tabular}
            }
    \footnotesize  \tiny Min: Minimum of Synchronization Degree, Max: Maximum of Synchronization Degree, NumSync: Number of Synchronization Episodes, FreqRatio: Average Heart-to-Respiration Rate Ratio.
\vspace{-0.6cm}
    \end{table}

\subsection{CRC Varies Between Rest and Recovery}
~\label{sec: CRC varies}



\textbf{CRC episodes are more frequent but less stable during exercise recovery, with a higher FreqRatio and a wider range of synchronization degree.} As shown in Table~\ref{table_1} and Fig~\ref{fig:box}, FreqRatio increased from $4.15$ at rest to $4.82$ during exercise recovery and the NumSync rose from $1.55$ to $3.15$. The range of synchronization degree also expanded, with $3.53\%-25.17\%$ at rest compared to $2.05\%-31.69\%$ during recovery. These results point to a more dynamic cardiorespiratory interaction during recovery.


\subsection{Respiratory Phase Synchronizes with Pulse Wave Dynamics}
~\label{sec:resp_phase}



\textbf{BVP decline corresponds to the respiratory phase of inhalation, while BVP acceleration corresponds to exhalation, reflecting the pattern between heartbeats and breathing cycles.} As shown in Table~\ref{table_2} and Fig~\ref{fig:bprsa}, the BPRSA-derived features—maximum respiratory amplitude (MRA) and slope at the anchor point (SAP)—exhibit significant differences across rest and exercise recovery (MRA: $0.02$ to $0.09$; SAP: $0.00$ to $0.10$ for BVP decline, and MRA: $0.02$ to $0.09$; SAP: $-0.01$ to $-0.10$ for BVP acceleration, all with $p<0.001$). This indicates that the heart-lung interaction is enhanced during exercise recovery when metabolic demands are elevated.


\subsection{rPPG Mirrors Oximeter-Based CRC Trends}
~\label{sec:remote_rppg}

\begin{figure}
    \centering
    \includegraphics[width=1\linewidth]{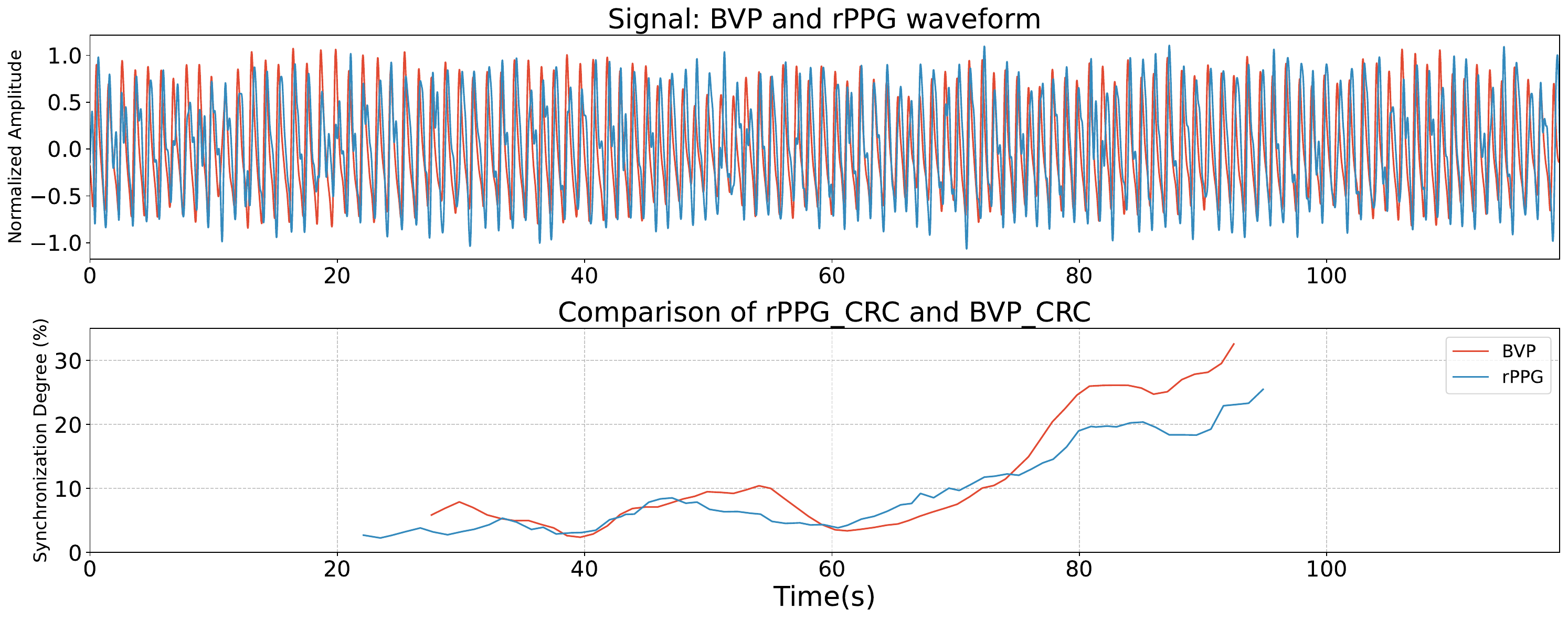}
        \vspace{-0.8cm}
    \caption{Comparison of cardiorespiratory coupling (CRC) curves derived from rPPG and BVP signals.}
    \label{fig:rppg_crc}
     \vspace{-0.3cm}
\end{figure}

\textbf{The CRC synchronization curves derived from rPPG measurements closely match those obtained from contact-based BVP signals, indicating the potential for non-contact CRC assessment.} The TN-rPPG model~\cite{wang2024plug}, trained on RLAP~\cite{wang2024camera}, was used to extract PPG waveforms from facial frames by performing inference on SUMS~\cite{liu2024summit} videos. As shown in Fig~\ref{fig:HR} and Tab~\ref{table_3}, the remote heart rate estimation achieves medical-grade accuracy (MAE $\leq 2$) and exhibits a high correlation of 0.99 with ground truth. As illustrated in Fig~\ref{fig:rppg_crc}, the predicted peaks of the rPPG and BVP waveforms align closely, and the calculated CRC curves exhibit a strong correlation (Pearson $r=0.956$). These results suggest that cameras can effectively capture cardiorespiratory dynamics in high altitudes, paving the way for non-invasive physiological monitoring.

\begin{table}
    \centering
    \setlength{\tabcolsep}{3pt}
    \caption{BPRSA (BVP->RESP)}
    \vspace{-0.3cm}
    \label{table_2}
        \resizebox{\columnwidth}{!}{
    \begin{tabular}{lcccc}\toprule
    Driven Signal & \multicolumn{2}{c}{BVP Decline} & \multicolumn{2}{c}{BVP Accelerate} \\\midrule
    State & MRA & SAP & MRA & SAP \\\midrule
    Stationary
        & $0.02 \pm 0.01$ & $0.00 \pm 0.01$ & $0.02 \pm 0.01$ & $-0.01 \pm 0.02$ \\
    Exercise Recovery
        & $0.09 \pm 0.05$ & $0.10 \pm 0.09$ & $0.09 \pm 0.06$ & $-0.10 \pm 0.09$ \\
    \bottomrule
    \end{tabular}}
    {\footnotesize  \tiny MRA: Maximum Respiratory Amplitude, SAP: Slope at Anchor Point.}
    \vspace{-0.6cm}
    \end{table}
    
\section{Limitation and Future Work}

While our findings demonstrate the feasibility of remote cardiorespiratory coupling (CRC) estimation, several limitations constrain the current study’s generalizability and scope. The participant sample was relatively limited in both size~\cite{tang2024camera} and diversity~\cite{tang2023mmpd}, which may restrict the robustness of observed CRC patterns across broader populations. In addition, although non-contact PPG assessment was achieved using a pretrained TN-rPPG model, respiratory signal extraction was not integrated within the same pipeline. To advance CRC as a potential clinical biomarker, future work will focus on refining non-contact algorithms to jointly estimate heart and respiratory signals with higher precision. We also envision incorporating foundation models and large language models (LLMs) to enable semantic-level understanding of multimodal physiological signals~\cite{tang2023alpha}. Such models may facilitate context-aware CRC estimation, enabling more scalable and generalizable health monitoring applications in both clinical and daily-life scenarios.

\section{Conclusion}

This study investigated cardiorespiratory coupling under high-altitude conditions, revealing significant differences ($p<0.05$) between rest and exercise recovery states. Using the Bivariate Phase-Rectified Signal Averaging (BPRSA) method, we analyzed the phase coupling relationship between blood volume pulse (BVP) and respiration (RESP) waveforms over short intervals (4~s). Furthermore, we explored the feasibility of remote cardiorespiratory assessment using non-contact rPPG signals, achieving accurate HR estimation (MAE $=$ 1.08 BPM) and CRC prediction (Pearson r $=$ 0.96). These results demonstrate the potential of remote video sensing for practical applications in physiological monitoring.

\begin{table}
    \centering
    \setlength{\tabcolsep}{3pt}
    \caption{Remote HR Estimation Results (TN-rPPG)}
    \vspace{-0.3cm}
    \label{table_3}
        \resizebox{\columnwidth}{!}{
    \begin{tabular}{lcccc}\toprule
    State&MAE$\downarrow$ & MAPE$\downarrow$ & RMSE$\downarrow$ & $\rho$ $\uparrow$  \\\midrule
    Stationary
        & $0.65 \pm 1.20$& $0.82 \pm 1.50$& $1.37 \pm 2.07$& 0.99\\
    Exercise Recovery
        & $1.51 \pm 1.54$& $1.49 \pm 1.50$& $2.16 \pm 2.50$&0.99\\
    All & $1.08\pm1.45$& $1.16\pm1.54$& $1.81\pm2.35$&0.99\\
    \bottomrule
    \end{tabular}}
    \footnotesize  \tiny MAE = Mean Absolute Error in HR estimation (Beats/Min), MAPE = Mean Percentage Error (\%), RMSE = Root Mean Square Error (Beats/Min), $\rho$ = Pearson Correlation in HR Prediction.
    \vspace{-0.5cm}
    \end{table}

\begin{acks}
This work is supported by the foundation of National Key Laboratory of Human Factors Engineering No. HFNKL2024W06, the National Natural Science Foundation of China No. 62366043 \& 62472244, Beijing Natural Science Foundation No.QY24248, the National Key R\&D Program of China No. 2024YFB4505500 \& 2024YFB4505503, the Tsinghua University Initiative Scientific Research Program No. 20257020004, and Qinghai University Research Ability Enhancement Project No. 2025KTSA05.
\end{acks}

\bibliographystyle{ACM-Reference-Format}
\bibliography{main}


\end{document}